\documentclass[conference]{IEEEtran}
\IEEEoverridecommandlockouts

\usepackage{cite}
\usepackage{amsmath,amssymb,amsfonts}
\usepackage{algorithmic}
\usepackage{graphicx}
\usepackage{textcomp}
\usepackage{xcolor}
\usepackage{tabularx}
\usepackage[numbers]{natbib}

\def\BibTeX{{\rm B\kern-.05em{\sc i\kern-.025em b}\kern-.08em
		T\kern-.1667em\lower.7ex\hbox{E}\kern-.125emX}}

\begin{document}
	
	\title{Beyond Correctness: How Novice Software Engineering Students Exercise Oversight of Generative AI
	}
	
	\author{\IEEEauthorblockN{Ali Shakiba}
		\IEEEauthorblockA{\textit{School of Electrical and Computer Engineering} \\
			\textit{The University of Sydney}\\
			Sydney, Australia \\
			ali.shakiba@sydney.edu.au}
		\and
		\IEEEauthorblockN{Thomas Chaffey}
		\IEEEauthorblockA{\textit{School of Electrical and Computer Engineering} \\
			\textit{The University of Sydney}\\
			Sydney, Australia \\
			thomas.chaffey@sydney.edu.au}
	}
	
	\maketitle
	
	\begin{abstract}
		Generative AI (GenAI) is shifting software engineering students' work from producing artefacts towards evaluating and supervising AI-generated outputs. However, little is known about what characteristics students identify in outputs when reporting interventions, or how they exercise oversight during authentic software engineering work. We analysed 962 student-AI interaction turns documented by 145 first-year software engineering students through structured AI Journals, including 172 student-declared interventions. Using a multi-dimensional coding framework, we characterised the output characteristics identified in these reported interventions, students' reasons for intervening, the actions they took, and how those actions were enacted.
		Fitness problems were identified in 74.4\% of interventions, compared with 16.9\% involving correctness problems, indicating that oversight more often concerned whether plausible AI outputs were suitable for the engineering task than whether they contained explicit errors. Among classifiable interventions, students most frequently modified generated outputs (51.0\%) or refined their prompts (28.2\%), with both response types commonly associated with engineering fit. Within-student analysis further showed that outputs perceived as containing minor flaws were associated with lower evaluation confidence than outputs perceived as perfect. Taken together, these findings characterise novice AI oversight as a process that extends beyond error detection to encompass both judgement of fitness for purpose and active intervention in AI-assisted engineering work.
	\end{abstract}
	
	
	\section{Introduction}\label{sec:intro}
	Generative AI (GenAI) tools have become a routine part of software engineering practice and education. Students increasingly use large language models (LLMs) to generate requirements specifications, user stories, interface designs, and code artefacts as part of their coursework \cite{denny2023,ghimire2024,sengul2024}. 
	As these systems produce increasingly fluent and plausible outputs, their use can shift part of students' work from \emph{creating} engineering artefacts from scratch towards \emph{evaluating}, adapting, and integrating AI-generated artefacts. This shift places students in a supervisory role, requiring them to decide whether to accept, refine, challenge, verify, or reject AI-generated outputs using their \emph{engineering judgement}.
	
	This supervisory role is particularly challenging for novice learners. Students may struggle to critically evaluate AI-generated outputs, especially when those outputs are fluent and structurally plausible \cite{jafari2026,misiejuk2026}. This creates a risk of cognitive offloading, whereby students rely on AI-generated outputs rather than independently interrogating them \cite{misiejuk2026}. 
	Huynh and Nguyen \cite{huynh2026}, for example, identify contrasting pathways: AI literacy can promote verification behaviour, whereas familiarity with AI and the fluency of its outputs can encourage uncritical adoption. The challenge, therefore, is not simply whether students use GenAI, but whether they can recognise when its outputs require scrutiny and exercise appropriate judgement when deciding how to respond.
	
	The educational capability underlying this supervisory role is \emph{evaluative judgement} (EJ): the capacity to assess the quality of one's own and others' work \cite{boud2000,tai2018}. EJ develops through sustained engagement with standards, comparison, and feedback, rather than through access to tools alone \cite{nicol2006,panadero2019,garg2025}. Bearman et al. \cite{bearman2024} extend this framing to AI-assisted learning, arguing that the ability to judge the fitness and trustworthiness of AI-generated artefacts is becoming a central dimension of EJ. In software engineering, this form of judgement can be operationalised through activities such as validation, testing, debugging, and comparison of AI-generated artefacts against domain requirements and constraints.
	
	Human-in-the-loop (HITL) frameworks complement this educational perspective by conceptualising human judgement as part of the operational workflow of an automated system \cite{kluter2026}. Parasuraman et al.'s Levels of Automation model \cite{parasuraman2000}, for example, highlights the risks to situation awareness and error detection as automation progresses from recommendation to execution. In AI-assisted coursework, a comparable decision arises whenever students determine whether and how to incorporate an AI-generated artefact into their engineering work. EJ and HITL thus illuminate complementary aspects of the same supervisory process: EJ concerns the student's capacity to judge the quality and fitness of an AI-generated artefact, while HITL foregrounds how that judgement is translated into actions such as accepting, modifying, verifying, or rejecting it.
	
	Prior process-level research has shown that students and developers engage with GenAI through prompting, inspection, verification, modification, and iteration \cite{kazemitabaar2023,barke2023,mozannar2024,prather2024,rahe2025}. However, less is known about the \emph{engineering judgement that underpins these behaviours}: what characteristics novice students identify in AI-generated artefacts when reporting interventions, why they judge those characteristics to require intervention, and how those judgements translate into particular responses. This gap is especially relevant in design-first, low-code software engineering curricula, where oversight extends beyond code correctness to include judgements about requirements, system scope, stakeholder needs, contextual fit, and design decisions \cite{arora2024}.
	
	We address this gap through a process-level empirical analysis of student oversight in an authentic first-year software engineering course. We analyse AI Journal entries \cite{shakiba_making_2026} submitted by 145 first-year students in a design-first, low-code course at the University of Sydney. The journals document students' prompts and AI-generated responses together with their self-reported evaluations, interventions, and reflective commentary. We analyse student-declared interventions across four dimensions: what characteristic of the output was identified when reporting an intervention, why the student intervened, what action they took, and how that action was enacted. We additionally examine how students' confidence in evaluating AI-generated outputs varies with their perceived correctness.
	\begin{description} 
		\item[\textbf{RQ1:}] What characteristics of AI-generated outputs do novice students identify when reporting interventions? 
		\item[\textbf{RQ2:}] How do novice software engineering students intervene in AI-generated outputs, and what operational mechanisms and engineering rationales characterise their interventions? 
		\item[\textbf{RQ3:}] How does students' confidence in evaluating AI-generated outputs vary with their perceived correctness of those outputs?
	\end{description}
	
	This paper makes three contributions: 
	\begin{enumerate} 
		\item an empirical characterisation of the AI-output characteristics identified in novice students' reported interventions, distinguishing problems of correctness from problems of fitness for the intended engineering task;
		\item a multi-dimensional framework linking characteristics identified in AI-generated outputs to students' engineering rationales, intervention actions, and operational mechanisms, together with empirical evidence of how these dimensions relate in students' reported oversight; and
		\item implications for software engineering education concerning how students can be supported to evaluate the fitness of AI-generated artefacts, exercise different forms of intervention, and make their supervisory judgement visible in assessment.
	\end{enumerate}
	
	The remainder of this paper is structured as follows.
	Section~\ref{sec:related} reviews related work on generative AI in software engineering
	education, evaluative judgement, and human-in-the-loop oversight frameworks.
	Section~\ref{sec:method} describes the research context, dataset, the AI Journal instrument, and the coding procedures.
	Section~\ref{sec:results} presents the results for each research question.
	Section~\ref{sec:discussion} discusses the findings and their implications for software
	engineering education practice.
	Section~\ref{sec:limitations} addresses limitations, and
	Section~\ref{sec:conclusion} concludes the paper.
	
	\section{Related Work}\label{sec:related}
	Empirical research has increasingly examined GenAI use through students' and developers' interaction processes, rather than focusing only on their attitudes towards these systems. Kazemitabaar et al. \cite{kazemitabaar2023} studied novice learners using an LLM-based code generator, examining prompting patterns alongside how students placed, verified, and manually modified generated code. They identified approaches ranging from single-prompt generation to hybrid AI-human coding. Related observational studies further demonstrate that working with code-generating systems involves ongoing human judgement. Barke et al. \cite{barke2023} distinguished between \emph{acceleration}, in which programmers use generated code to implement an already-understood solution, and \emph{exploration}, in which they use the model to investigate possible approaches. Mozannar et al. \cite{mozannar2024} developed a complementary taxonomy of programmer activities around GitHub Copilot, capturing the activities and associated costs involved in prompting, inspecting, and working with code suggestions.
	
	Research with novice programmers further indicates that access to GenAI does not necessarily result in effective oversight. Prather et al. \cite{prather2024} found that GenAI accelerated problem-solving for some novices while compounding existing metacognitive difficulties for others. Rahe and Maalej \cite{rahe2025} similarly observed programming students using ChatGPT both to seek conceptual knowledge and generate solutions, with some entering repeated cycles of submitting generated code and asking the model to revise unsuccessful outputs. In response to these challenges, pedagogical approaches have begun to emphasise capabilities such as prompt construction, code reading, and evaluation. For example, Denny et al.'s \emph{Prompt Problems} train students to formulate prompts that satisfy programming specifications \cite{denny2023}. Together, these studies show that novice interactions with GenAI in programming contexts involve iterative prompting, inspection, verification, and modification, while revealing substantial variation in how critically students engage in these activities.
	
	Much of this process-level evidence, however, concerns code-generation or programming tasks, where output quality can often be evaluated against functional behaviour or predefined specifications. GenAI is also increasingly being applied to upstream software engineering activities, including requirements elicitation, specification, analysis, and validation \cite{arora2024}. Quality in these activities is multidimensional: an artefact may be internally plausible yet still require judgement about its completeness, consistency, scope, alignment with stakeholder needs, and fit with project constraints \cite{arora2024,mircea2026}. The oversight problem therefore extends beyond determining whether generated content is technically or factually correct to assessing whether it is suitable for the engineering context in which it will be used.
	
	Beyond interaction strategies, research on verification raises questions about the quality and consistency of students' evaluative practices. Ardic et al. \cite{ardic2026} observed that students working with AI on software testing tasks inspected outputs and refined their prompts following weak results, although manual validation practices varied considerably across individuals and task types. Cognitive offloading, whereby AI-generated artefacts substitute for independent reasoning, has also emerged as a longitudinal concern, with students' debugging and critical-reasoning capabilities weakening across course cycles \cite{misiejuk2026}. Together, these findings indicate that opportunities to inspect or modify AI-generated outputs do not necessarily translate into effective oversight.
	
	Students' trust in AI outputs may also be miscalibrated. Huynh and Nguyen \cite{huynh2026} identify two partly independent pathways shaping reliance: one driven by familiarity, perceived usefulness, and self-efficacy, which increases willingness to adopt AI, and another driven by AI literacy and habitual verification, which strengthens auditing capacity. Students may therefore be highly familiar with a tool while remaining poorly equipped to distinguish reliable from unreliable outputs. Pitts et al. \cite{pitts2027} further show that neither self-reported trust nor expressed scepticism reliably predicts auditing behaviour in practice. Taken together with process-level studies of prompting, verification, and modification, this evidence indicates that understanding responsible GenAI use requires attention to both students' evaluative dispositions and the actions through which their evaluations are enacted.
	
	EJ provides an educational lens through which to understand this capability. EJ develops through sustained engagement with standards, comparison, and criterion-referenced feedback, rather than through access to tools alone \cite{tai2018,nicol2006,panadero2019}. In technical education, EJ involves assessing work against relevant quality criteria, identifying deficiencies, and making justified decisions about its adequacy \cite{tai2018}. Bearman et al. \cite{bearman2024} extend this framing to GenAI, arguing that judging the fitness and trustworthiness of AI-generated artefacts is becoming a central dimension of EJ in higher education. Crucially, EJ in AI-assisted contexts entails not only recognising explicit errors but also assessing whether an apparently plausible output is suitable for its intended context and purpose. Such judgement may not be visible in the final artefact or explicitly rewarded by existing assessment designs \cite{bearman2024}.
	
	HITL perspectives complement EJ by locating this judgement within AI-mediated workflows. Rather than treating human involvement as the passive monitoring of an otherwise automated process, HITL emphasises meaningful opportunities for people to evaluate, redirect, modify, override, or reject system outputs \cite{kluter2026}. Parasuraman et al.'s Levels of Automation model provides a foundational account of how responsibility can be distributed between humans and automated systems \cite{parasuraman2000}. More recent reviews of HITL in applied AI emphasise that assigning a person a nominal supervisory role does not, by itself, constitute meaningful oversight \cite{kluter2026}. This distinction is increasingly important as AI-assisted and agentic workflows distribute software engineering activities across human and automated actors. Effective human involvement depends on whether people retain both the capacity and opportunity to exercise judgement when generated artefacts require intervention.
	
	Existing process-level research therefore provides substantial evidence about \emph{how} students and programmers interact with GenAI, including how they prompt, inspect, verify, modify, and iterate on generated outputs. Less well understood is the engineering judgement underlying these behaviours: which characteristics students identify in AI-generated artefacts when reporting interventions, why students judge those characteristics to require intervention, and how their judgements translate into particular responses. This gap is especially important beyond code-generation tasks, where an output may be technically plausible yet insufficiently scoped, contextualised, or aligned with stakeholder and project requirements. We address this gap through a semester-long analysis of student-reported oversight across four related dimensions: the characteristics identified in AI-generated outputs, students' engineering rationales, their intervention actions, and the operational mechanisms through which those actions were enacted in authentic software engineering coursework.
	
	\section{Methodology}\label{sec:method}
	\subsection{Research Context and the AI Journal} 
	This study was conducted in ELEC1005: Introduction to Software Engineering, a first-year undergraduate unit at the University of Sydney in Semester 1, 2026. The unit introduces core Software Development Lifecycle (SDLC) concepts, including requirements engineering, stakeholder analysis, user stories, interface design, and prototyping. It follows a design-first, low-code approach in which students use tools such as Microsoft PowerApps and Figma rather than developing software primarily through conventional source-code programming. Because the course emphasised requirements interpretation, system design, prototyping, and the justification of engineering decisions rather than algorithmic implementation, students frequently used GenAI for activities where contextual suitability and system constraints were particularly salient.
	
	As part of the coursework, students completed structured AI Journals throughout the semester \cite{shakiba_making_2026}. Journal completion was a hurdle requirement rather than a graded component, thereby reducing, but not eliminating, students’ incentive to optimise their reflective responses for marks. Each journal documented a student's use of a GenAI tool for a coursework task and contained one or more interaction turns. For each turn, students recorded their prompt and the corresponding AI-generated response, the purpose of the interaction, their assessment of the output's accuracy, the method used to verify the output, and their confidence in evaluating its accuracy on a five-point scale. When students assessed an output as containing \textsc{Minor Flaws} or \textsc{Major Hallucination Errors}, they were also required to describe any intervention they made. At the journal level, students reflected on how their interaction process evolved, critical moments in that process, perceived skill dependency, the utility of the AI, and the interaction overall. The resulting records therefore combine documented student-AI interaction turns with students' self-reported evaluations, interventions, and reflections.
	
	\subsection{Dataset and Unit of Analysis}    
	The dataset comprises 962 student-AI interaction turns declared by 145 students across ten weeks. Each journal entry represents one or more documented student-AI interactions and contains the student's prompts, the AI-generated responses, and the associated reflective fields. 
	
	For RQ1 and RQ2, the primary unit of analysis is the \emph{student-declared intervention}. 
	At each interaction turn, students first evaluated the correctness of the AI-generated output as \emph{Perfect}, with \emph{Minor Flaws}, or with \emph{Major Hallucinations}. When an output was evaluated as containing minor flaws or major hallucinations, students were required to describe the human intervention they made in response and how they carried it out. These intervention declarations, together with the corresponding prompt, AI-generated response, and surrounding reflective fields, constitute the intervention dataset analysed for RQ1 and RQ2.
	
	For RQ3, we consider students' self-reported confidence in evaluating the accuracy of AI-generated outputs, recorded on a five-point scale for each interaction turn, in relation to their perceived correctness of the corresponding output.
	
	\subsection{Human Oversight Coding Framework}\label{sec:method:framework} 
	We developed the coding framework iteratively from the complete set of student-declared interventions. The first author initially reviewed all intervention records to identify recurring characteristics in the factors that prompted students to intervene, their reasons for intervening, the actions they took, and how those actions were enacted. Based on this review, an initial set of codes, definitions, and decision rules was developed. The full intervention dataset was then reviewed a second time using this provisional framework, with the categories and their definitions refined to resolve ambiguities and improve conceptual separation. 
	
	Rather than treating oversight as a binary distinction between acceptance and rejection, the framework captures four complementary dimensions of the intervention process: (1) the characteristic of the AI-generated output that prompted the intervention (\emph{Trigger}, TR); (2) the student's underlying engineering motivation for intervening (\emph{Intent and Rationale}, IR); (3) the action taken in response (\emph{Intervention Type}, IT); and (4) the means by which the intervention was implemented (\emph{Operational Mechanism}, OM). Together, these dimensions characterise an intervention in terms of what prompted it (TR), why the student acted (IR), what action they took (IT), and how that action was enacted (OM).
	
	\begin{table}[ht]
		\caption{Human Oversight Coding Framework: Trigger (TR), Intent and Rationale (IR), Intervention Types (IT), and Operational Mechanisms (OM) codes.} 
		\label{tab:coding-framework} 
		\centering 
		\scriptsize 
		\setlength{\tabcolsep}{4pt} 
		\begin{tabular}{p{.04\columnwidth} p{.26\columnwidth} p{.55\columnwidth}} 
			\hline \textbf{Code} & \textbf{Label} & \textbf{Definition} \\ \hline
			\multicolumn{3}{c}{\textit{Intervention Trigger (TR) - what prompted the student to intervene}} \\ \hline 
			TR1 & Correctness Problem &
			The output contains a perceived factual, logical, technical, or structural/formatting error that requires correction. \\
			TR2 & Fitness Problem  &
			The output is not considered fit for the intended task, project, or context, for example because it is overly generic, complex, incomplete, poorly focused, contextually misaligned, or inconsistent with the intended system scope. \\
			TR3 & No Identifiable Failure &
			No specific deficiency in the AI-generated output can be identified from the student's account, despite a reported response or intervention. \\ \hline
			\multicolumn{3}{c}{\textit{Intent and Rationale (IR) - why the student acted}} \\ \hline 
			IR1 & Engineering Fit &
			Ensuring that the output is feasible, practical, and compatible with
			the project, domain, or system constraints. \\
			IR2 & Comprehensibility &
			Making the output clearer, simpler, or easier for the student to
			understand and work with. \\
			IR3 & Assessment Alignment &
			Aligning the output with assessment requirements, learning materials,
			or marking criteria.
			\\
			\hline \multicolumn{3}{c}{\textit{Intervention Types (IT) - what the student did}} \\ 
			\hline IT1 & Output Modification & Student directly modifies, corrects, restructures, or contextualises the generated output. \\ 
			IT2 & Prompt Refinement & Student changes the instructions/context supplied to the AI to obtain a more suitable output. \\ 
			IT3 & Clarification Seeking & Student asks the AI to explain, elaborate, or clarify its output without modifying it. \\ 
			IT4 & Rejection & Student rejects, discards, or chooses not to use the generated output. \\ 
			IT5 & Insufficient Intervention Description & The student's intervention declaration was incomplete, unclear, or insufficiently relevant to determine the action taken, and no intervention type could be reliably inferred from the surrounding interaction context. \\ \hline
			\multicolumn{3}{c}{\textit{Operational Mechanisms (OM) - how the student implemented it}} \\ \hline 
			OM1 & Evidence Grounding & Providing external artefacts, references, documentation, or other evidence to ground the interaction. \\ 
			OM2 & Specification Adjustment & Adding, revising, or making explicit requirements, constraints, or specifications to guide the AI response. \\ 
			OM3 & Output Transformation & Selecting, simplifying, contextualising, restructuring, or reformatting generated content to make it suitable for the intended use. \\ 
			OM4 & Error Diagnosis & Identifying or reasoning about a specific factual, logical, technical, or other problem in the generated output. \\ 
			\hline 
		\end{tabular} 
	\end{table} 
	
	IT codes were assigned only where the student's declaration or surrounding interaction context provided sufficient evidence for the action taken. Where a student's account described multiple actions, the IT code was assigned based on the primary action taken in response to the AI-generated output. Where an intervention was declared but the available evidence did not support assignment to any of IT1-IT4, the case was coded as IT5.
	
	The Intervention Trigger (TR) captures the characteristic of the AI-generated output that prompted the student’s reported intervention. TR reflects the student’s characterisation of the output and their reason for responding to it, rather than an independent judgement by the researchers about whether the output was objectively incorrect. The student’s prompt, the AI-generated response, the declared intervention, and the associated reflective fields were considered together when assigning one or more of three TR codes: \textsc{Correctness Problem} (TR1), \textsc{Fitness Problem} (TR2), or \textsc{No Identifiable Failure} (TR3).
	
	TR1 and TR2 distinguish deficiencies in the validity of the generated artefact from those in its suitability for the intended engineering context. For example, a technically invalid claim or an internally inconsistent requirement was coded as a correctness problem (TR1), whereas an otherwise plausible suggestion that conflicted with the project scope, stakeholder needs, platform constraints, or assessment requirements was coded as a fitness problem (TR2). Where evidence supported both conditions, both codes were assigned.
	
	\subsection{Coding Procedure} \label{sec:method:coding} 
	Using the framework described in Section \ref{sec:method:framework}, both authors independently coded the intervention events and then compared their assignments. For the single-label IT dimension, inter-rater agreement was $90.6\%$ (Cohen's $\kappa=0.865$; Gwet's AC1$=0.886$). For the multi-label dimensions, agreement was assessed using exact-set agreement and Jaccard similarity. Exact-set agreement was $94.7\%$ for OM (mean Jaccard$=0.947$), $91.3\%$ for IR (mean Jaccard$=0.920$), and $92.1\%$ for TR (mean Jaccard$=0.930$). After completing the independent coding, the authors resolved all disagreements through discussion, referring to the coding definitions and decision rules to reach consensus. All subsequent analyses were conducted using this reconciled human-coded dataset.
	
	\subsection{Data Analysis}\label{sec:method:analysis}
	For RQ1, we examined the prevalence of intervention triggers across student-declared interventions. Because multiple triggers could be identified for a single intervention, TR codes were treated as multi-label, and prevalence was calculated as the proportion of interventions in which each trigger was identified. We also examined the underlying journal records to contextualise the observed trigger patterns using representative examples.
	
	For RQ2, we first examined the distribution of primary intervention types among student-declared interventions. Cases in which the action taken could not be determined were retained as \textsc{Insufficient Intervention Description} (IT5) but excluded from the calculation of the relative distribution of the four substantive intervention types. An intervention could receive multiple codes within IR and OM dimensions when supported by the student's account and surrounding interaction context. Where the available evidence was insufficient to assign any code within a dimension, no code was assigned. We then examined the rationales and operational mechanisms associated with each intervention type using pairwise co-occurrence frequencies for IT-IR and IT-OM. Because IR and OM codes were multi-label, co-occurrences were calculated independently for each dimension and normalised within intervention type. 
	
	For RQ3, we examined the relationship between students' perceived output correctness and their confidence in evaluating the accuracy of AI-generated outputs. For descriptive visualisation, confidence scores were mean-centred within student as $\Delta C_{s,i}=C_{s,i}-\mu_s$, where $C_{s,i}$ is the confidence reported by student $s$ for interaction turn $i$, and $\mu_s$ is that student's mean confidence across all interaction turns.
	
	The inferential analyses were restricted to interactions evaluated as \textsc{Perfect} or \textsc{Minor Flaws}. The five interactions evaluated as \textsc{Major Hallucination Errors} were retained in the descriptive analysis but excluded from inferential comparisons because there were too few observations to support reliable estimation. Perceived correctness was represented by a binary indicator, $M_{s,i}$, coded as $0$ for \textsc{Perfect} and $1$ for \textsc{Minor Flaws}.
	
	Because individual students contributed repeated interaction turns, we fitted a linear mixed-effects model \cite{lindstrom1988newton,seabold2010statsmodels} with a student-level random intercept. To distinguish within-student associations from between-student differences, we decomposed perceived correctness into within-student and between-student components:
	\begin{equation}
		\label{eq:lin:mixed:model}
		C_{s,i}=\beta_0+\beta_W\left(M_{s,i}-\overline{M}_s\right)+\beta_B \overline{M}_s + u_s + \varepsilon_{s,i},
	\end{equation}
	where $\overline{M}_s$ is the proportion of interactions of student $s$ evaluated as \textsc{Minor Flaws}. The coefficient $\beta_W$ therefore estimates the within-student association between perceiving an output as having minor flaws and evaluation confidence, independently of between-student differences in the general tendency to evaluate outputs as flawed.
	
	Because confidence was recorded on an ordinal five-point scale, we conducted a sensitivity analysis using proportional-odds ordinal logistic regression \cite{seabold2010statsmodels,mccullagh1980regression} with the same within-student and between-student decomposition and standard errors clustered by student. We report the mixed-model estimate of $\beta_W$ with its $95\%$ confidence interval (CI), together with the corresponding odds ratio (OR) from the ordinal model. 
	
	\section{Results}\label{sec:results}
	\subsection{RQ1: Characteristics of AI Outputs Identified in Reported Interventions}\label{sec:results:rq1}
	Across the 172 student-declared interventions identified among 962 declared interaction turns, we identified three categories of intervention triggers: \textsc{Correctness Problem} (TR1), \textsc{Fitness Problem} (TR2), and \textsc{No Identifiable Failure} (TR3). \textsc{Fitness Problem} was by far the most prevalent trigger, occurring in 128 interventions (74.4\%), compared with 29 interventions (16.9\%) involving a \textsc{Correctness Problem}. In 26 cases (15.1\%), students declared an intervention but their account did not identify a specific deficiency that could be classified as either a correctness or fitness problem; these cases were coded as \textsc{No Identifiable Failure} (TR3). TR1 and TR2 were not mutually exclusive and could co-occur where an output exhibited both correctness and fitness problems. TR3 was assigned only where neither TR1 nor TR2 could be identified. Percentages therefore do not sum to 100\%.
	
	\textsc{Fitness Problem} captured outputs that were not necessarily incorrect but were unsuitable for the student's immediate engineering task. These included outputs that were overly generic, incomplete, unnecessarily complex, insufficiently contextualised, or misaligned with the intended system scope. For example, a student designing a medical-device workflow found that the generated response lacked sufficient project-specific context and subsequently provided additional information about the device and task.
	
	Overall, fitness problems were identified substantially more often than correctness problems among student-declared interventions.
	
	\subsection{RQ2: Student Oversight Behaviours}\label{sec:results:rq2}
\begin{figure}[h]
		\centering
		\includegraphics[width=\linewidth]{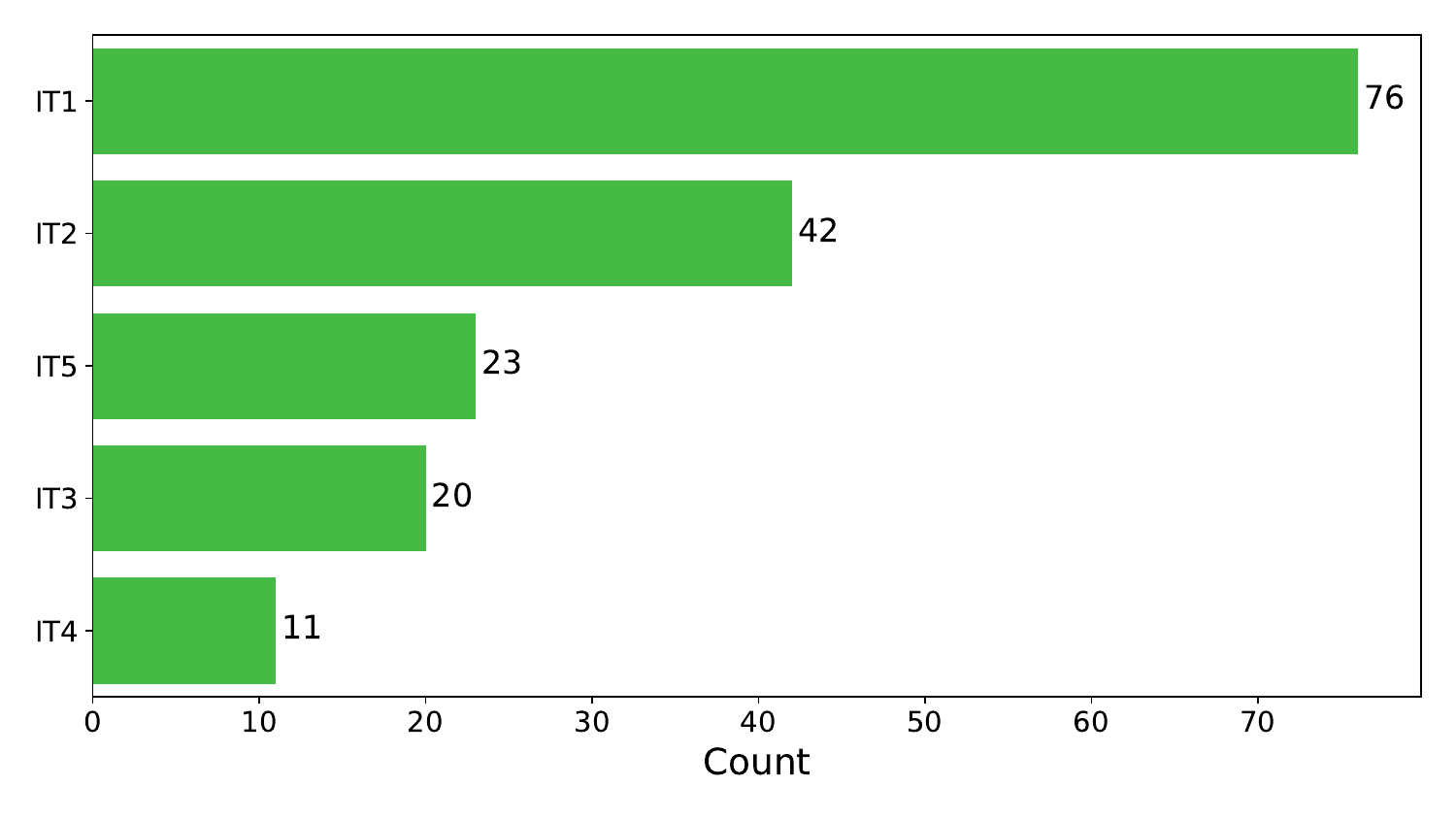}
		\caption{Distribution of intervention types across student-declared
			interventions ($N=172$). IT1-IT4 represent substantive intervention
			types; IT5 represents insufficient intervention descriptions.}
		\label{fig:it-distribution}
	\end{figure}

	RQ2 examines how students acted once they had decided to intervene, the engineering rationales underlying those actions, and the mechanisms through which the interventions were enacted. Of $172$ student-declared interventions, $23$ ($13.4\%$) were coded as \textsc{Insufficient Intervention Description} (IT5), as the intervention description and surrounding interaction context did not provide sufficient evidence to determine the action taken. Of the remaining $149$ interventions, \textsc{Output Modification} (IT1) was the most frequent ($51.0\%$, $n=76$), followed by \textsc{Prompt Refinement} (IT2; $28.2\%$, $n=42$), \textsc{Clarification Seeking} (IT3; $13.4\%$, $n=20$), and \textsc{Rejection} (IT4; $7.4\%$, $n=11$) (Figure \ref{fig:it-distribution}).
	
	The relationship between rationale and response type revealed distinct patterns in how students exercised oversight. \textsc{Engineering Fit} (IR1) was associated with $63.2\%$ ($48/76$) of \textsc{Output Modification} (IT1) interventions and $69.0\%$ ($29/42$) of \textsc{Prompt Refinement} (IT2) interventions. By contrast, \textsc{Clarification Seeking} (IT3) reflected a different pattern: \textsc{Comprehensibility} (IR2) was identified in $85.0\%$ ($17/20$) of these interventions.
	
	Operational mechanisms also differed across intervention types. \textsc{Output Transformation} (OM3) was the most frequently identified mechanism overall ($n=102$, $59.3\%$) and characterised $88.2\%$ ($67/76$) of \textsc{Output Modification} (IT1) interventions. \textsc{Prompt Refinement} (IT2) involved both \textsc{Output Transformation} (OM3) ($52.4\%$, $22/42$) and \textsc{Specification Adjustment} (OM2) ($40.5\%$, $17/42$), whereas \textsc{Clarification Seeking} (IT3) was most frequently associated with \textsc{Specification Adjustment} (OM2) ($75.0\%$, $15/20$). \textsc{Evidence Grounding} (OM1) and \textsc{Error Diagnosis} (OM4) were comparatively uncommon overall, occurring in eight ($4.7\%$) and twelve ($6.9\%$) interventions, respectively. Because interventions could involve multiple operational mechanisms, these percentages are not mutually exclusive.
	
	\begin{table}[t]
		\centering
		\caption{Distribution of intent and rationale (IR) and operational mechanism (OM) codes across student-declared interventions ($N=172$). IR and OM are multi-label dimensions; an intervention could receive multiple codes or no code when the available evidence was insufficient. Percentages therefore do not sum to 100\%.}
		\label{tab:ir-om-prevalence}
		\small
		\begin{tabular}{llrr}
			\hline
			& \textbf{Category} & \textbf{n} & \textbf{\%} \\
			\hline
			IR1 & Engineering Fit & 89 & 51.7 \\
			IR2 & Comprehensibility & 47 & 27.3 \\
			IR3 & Assessment Alignment & 46 & 26.7 \\
			\hline
			OM3 & Output Transformation & 102 & 59.3 \\
			OM2 & Specification Adjustment & 37 & 21.5 \\
			OM4 & Error Diagnosis & 12 & 6.9 \\
			OM1 & Evidence Grounding & 8 & 4.6 \\
			\hline
		\end{tabular}
	\end{table}
	
\begin{figure*}
		\centering
		\includegraphics[width=.7\textwidth]{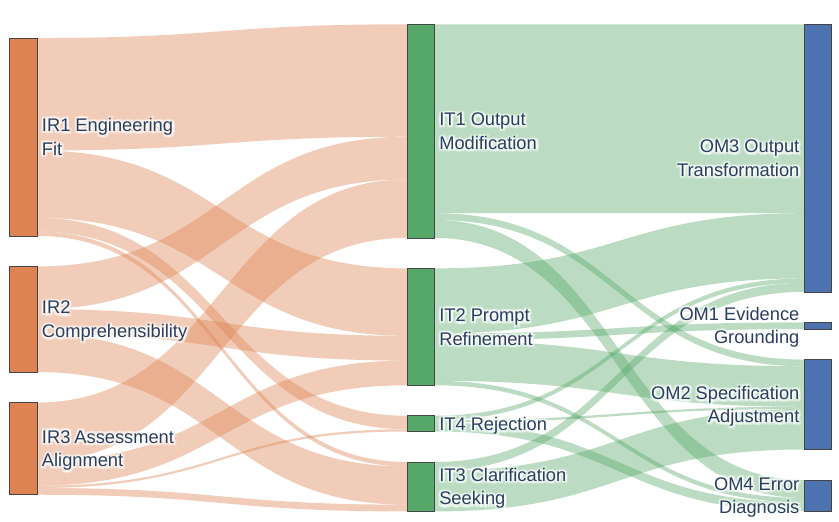}
		\caption{Pairwise associations among students' intent and rationale (IR), intervention type (IT), and operational mechanism (OM). Flow widths represent co-occurrence frequencies between IR and IT (left) and between IT and OM (right). IR and OM are multi-label dimensions; flows therefore represent pairwise associations rather than complete or causal sequences. IT5 cases are omitted because a substantive intervention type could not be determined.}
		\label{fig:oversight_relationships}
	\end{figure*}

	Figure \ref{fig:oversight_relationships} synthesises these pairwise relationships across the three dimensions, showing how intervention rationales were associated with response types and how those responses were operationalised. Overall, students most frequently exercised oversight by directly modifying AI-generated outputs or refining their prompts. Both response types were predominantly associated with \textsc{Engineering Fit}, whereas \textsc{Clarification Seeking} showed a distinct orientation towards \textsc{Comprehensibility}.
	
	\subsection{RQ3: Confidence and Perceived Output Correctness}\label{sec:results:rq3}
\begin{figure}[h]
		\centering
		\includegraphics[width=\linewidth]{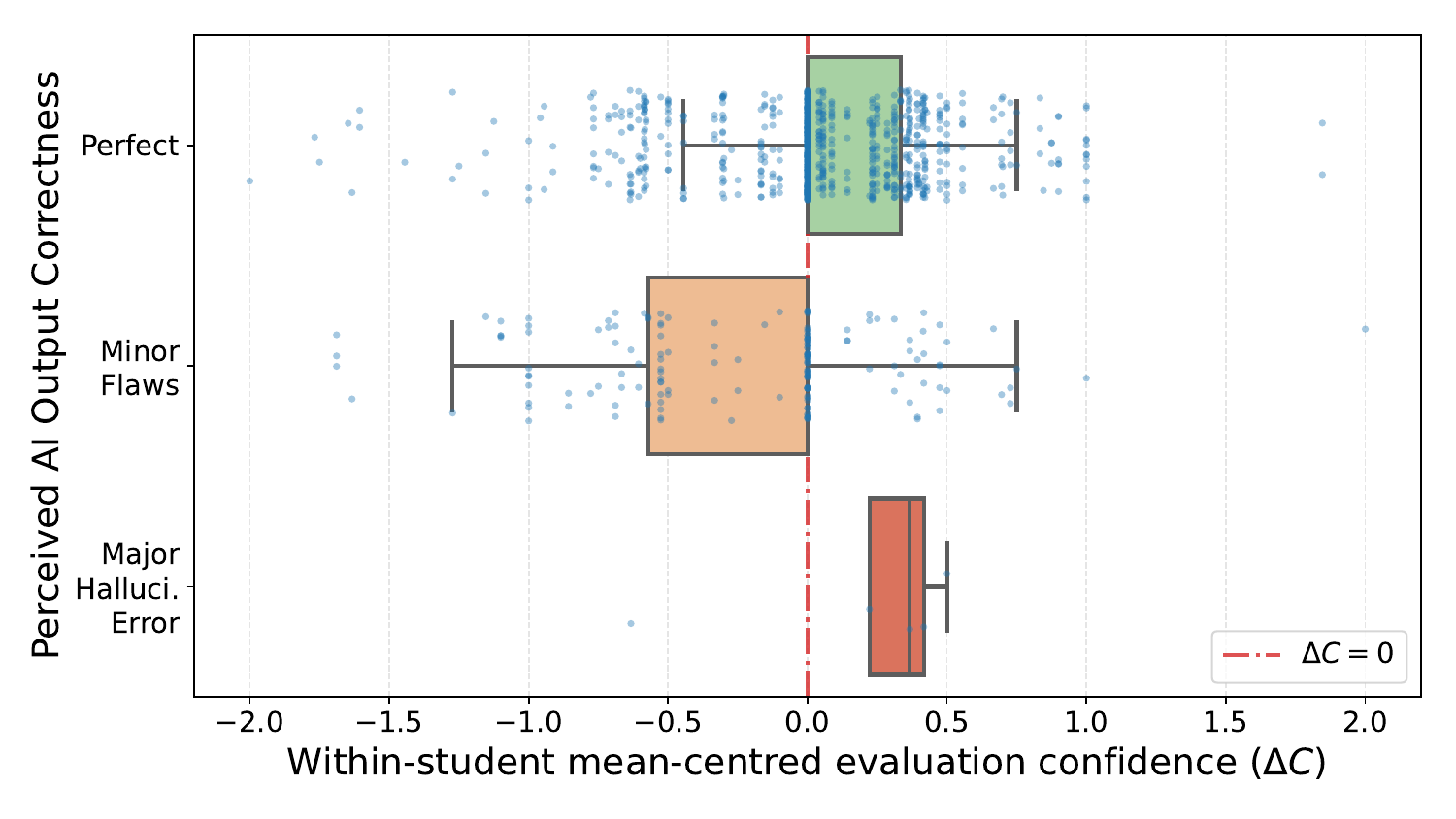}
		\caption{Distribution of within-student mean-centred evaluation confidence ($\Delta C$) by student-reported output correctness. $\Delta C=0$ represents a student's mean confidence across their interaction turns; negative and positive values indicate lower- and higher-than-usual evaluation confidence, respectively.}
		\label{fig:accuracy_confidence}
	\end{figure}

	Across the $962$ interaction turns, students evaluated most AI-generated outputs as \textsc{Perfect} ($n=790$, $82.2\%$), followed by \textsc{Minor Flaws} ($n=167$, $17.3\%$). $25$ students contributed observations in both the \textsc{Perfect} and \textsc{Minor Flaws} categories. Only five interactions ($0.5\%$) were evaluated as \textsc{Major Hallucination Errors}. Figure~\ref{fig:accuracy_confidence} shows the distribution of within-student mean-centred confidence, $\Delta C$, across the three perceived correctness categories.
	
	Evaluation confidence was lower when the same student perceived an output as containing \textsc{Minor Flaws} rather than as \textsc{Perfect}. After separating within- and between-student variation, the mixed-effects model estimated a $0.43$-point reduction in confidence for interactions perceived as containing minor flaws ($\beta_W=-0.434$, $95\%$ CI: $[-0.543,-0.325]$, $p<0.001$). A between-student association was also observed, students who more frequently evaluated outputs as containing minor flaws tended to report lower evaluation confidence overall ($\beta_B=-0.721$, $95\%$ CI: $[-1.095,-0.348]$, $p<0.001$). The estimated student-level random-intercept variance was $0.266$, with residual variance of $0.243$. The within-student association was robust to treating the five-point confidence measure as ordinal in a proportional-odds model (OR $=0.289$, $95\%$ CI: $[0.165,0.507]$, $p<0.001$).
	
	The five interactions evaluated as \textsc{Major Hallucination Errors} were reported by four students and had a mean raw confidence score of $4.60$ out of $5$. Given the small number of observations, these cases were retained for descriptive presentation only.
	
	\section{Discussion}\label{sec:discussion}
	\subsection{Oversight as Judgement of Fitness}
	A central finding is that novice oversight was concerned far more often with the \emph{fitness} of AI-generated artefacts than with their \emph{correctness}: fitness problems were identified in $74.4\%$ of student-declared interventions, compared with $16.9\%$ involving correctness problems. This distinction extends the framing of AI oversight beyond error detection. In software engineering, an artefact can be technically plausible yet unsuitable for its intended use because it does not adequately reflect project scope, stakeholder needs, platform constraints, or implementation context. Effective oversight therefore requires students to ask not only \emph{``Is this output correct?''}, but also \emph{``Is this output appropriate for what we are trying to build?''}
	
	This finding connects EJ and HITL perspectives in AI-assisted software engineering. EJ concerns judging work against relevant standards and contextual requirements \cite{tai2018,bearman2024}, while HITL foregrounds the human role in deciding whether to accept, modify, or override automated output \cite{parasuraman2000,kluter2026}. Our findings suggest that developing this supervisory capability requires more than teaching students to detect hallucinations or verify factual correctness: students also need to judge whether AI-generated artefacts are sufficiently specific, appropriately scoped, and contextually suitable for the engineering task at hand. We therefore identify \emph{fitness for purpose} as an important dimension of evaluative judgement in AI-assisted software engineering.
	
	\subsection{Oversight as Active Engineering Agency}
	The results in Section \ref{sec:results:rq2} show that novice oversight was not limited to verifying AI output or repeatedly prompting the model. Rather, students exercised agency at different points in the human-AI workflow: they directly modified generated artefacts, redirected the AI through prompt refinement, sought clarification, or rejected outputs altogether. Most classifiable interventions involved either \emph{Output Modification} (IT1; $51.0\%$) or \emph{Prompt Refinement} (IT2; $28.2\%$), both of which were most frequently associated with \emph{Engineering Fit} (IR1). This pattern suggests that students were not simply deciding whether to accept an AI response; instead, they actively shaped either the generated artefact or the conditions under which the AI produced it to better satisfy engineering requirements.
	
	The distinction between these response types also illustrates different forms of supervisory agency. \textsc{Output Modification} (IT1) places the student directly in control of the artefact, whereas \textsc{Prompt Refinement} (IT2) retains the AI as the producer while the student adjusts the constraints and context that guide its output. \textsc{Clarification Seeking} (IT3) serves a different function: its association with \textsc{Comprehensibility} (IR2) suggests that students also intervened to improve their own understanding before deciding how to proceed. Together, these behaviours illustrate a richer form of HITL oversight than a simple accept-reject decision \cite{parasuraman2000,kluter2026}: novice oversight can involve modifying the artefact, redirecting the automated system, or interrogating it to support further judgement. For software engineering education, this suggests that supervisory competence should encompass a range of intervention strategies, rather than treating oversight as a binary decision to accept or reject AI-generated output.
	
	\subsection{Evaluative Confidence and Imperfect Outputs}
	Results in Section \ref{sec:results:rq3} show that students' evaluation confidence was sensitive to perceived imperfections in AI-generated output. Within students, outputs perceived as containing minor flaws were associated with significantly lower evaluation confidence than outputs perceived as perfect, and this relationship remained robust when confidence was modelled as an ordinal measure. Students who more frequently evaluated outputs as containing minor flaws also tended to report lower evaluation confidence overall, indicating that the association was present both within and between students. These findings suggest that recognising a problem does not necessarily imply confidence in judging its extent or consequences; rather, subtle imperfections may introduce evaluative uncertainty.
	
	Together with the results in Section \ref{sec:results:rq1}, this finding highlights an important challenge for AI-assisted software engineering education. Most declared interventions involved fitness rather than correctness problems, while perceived minor flaws were associated with lower evaluation confidence. Supervisory competence therefore requires not only recognising that an AI-generated artefact may require intervention, but also developing sufficient evaluative judgement to determine why it is unsuitable and how it should be addressed. The five major-hallucination cases were too few to determine whether conspicuous failures elicit a different confidence response. In other words, the challenge is not simply noticing that something is wrong. Students may recognise that an output is imperfect while remaining uncertain about how confidently they can evaluate it.
	
	\subsection{Implications for Software Engineering Education}
	Our findings suggest that AI oversight should be treated as an explicit software engineering competency rather than an implicit expectation of students using GenAI. First, instruction should extend beyond factual verification to develop students' ability to judge \emph{fitness for purpose}, including whether generated artefacts are appropriately scoped, sufficiently contextualised, and compatible with system and stakeholder constraints. Second, students should develop a range of supervisory strategies. Effective intervention may involve directly adapting an artefact, refining the conditions provided to the AI, seeking clarification to improve understanding, or rejecting an output, rather than relying predominantly on repeated prompting.
	
	Finally, assessment should make this supervisory process visible. Evaluating only the final artefact provides limited evidence of whether students critically evaluated AI-generated content or accepted it uncritically. Process-oriented activities that require students to document what prompted an intervention, why they intervened, and how they responded can provide evidence of the evaluative judgement underlying AI-assisted engineering work. Such evidence can support feedback and assessment focused not simply on whether students use GenAI, but on the quality of the engineering judgement they exercise when using it.
	
	\section{Limitations}\label{sec:limitations}
	Our findings should be interpreted in light of several limitations. First, the AI Journal captures \emph{student-declared} oversight rather than independently observed behaviour \cite{shakiba_making_2026}. Students may omit interventions, provide incomplete descriptions, or reconstruct their reasoning when documenting an interaction. Moreover, intervention reporting was required only when students evaluated an output as containing \textsc{Minor Flaws} or \textsc{Major Hallucination Errors}, which may have excluded adaptations to outputs rated \textsc{Perfect}, such as modifications for contextual fitness rather than correctness. The additional reporting required after a non-perfect evaluation may also have influenced response choices. Furthermore, because the AI Journal explicitly prompted students to consider verification and evaluation confidence, the observed behaviours occurred within a scaffolded reflective activity and may not reflect students' GenAI use outside of the AI Journal activity. Consequently, the absence of a reported intervention cannot be interpreted as evidence that no intervention occurred. Similarly, the coded triggers, rationales, and actions represent only what could be established from students' accounts and the surrounding interaction context. The presence of intervention declarations with insufficient information for substantive classification (IT5) illustrates this limitation.
	
	Second, the coding framework necessarily involves interpretive judgement. Although coding considered the student's prompt, the AI-generated response, intervention declaration, and reflective fields together, distinctions between categories such as engineering fit, comprehensibility, and assessment alignment were not always explicit in students' accounts. We mitigated this threat through independent coding using explicit decision rules and assessed inter-rater reliability using Cohen's $\kappa$ \cite{cohen1960coefficient} and Gwet's AC1 \cite{gwet2008computing} before resolving disagreements through discussion (Section~\ref{sec:method:coding}). Although the observed high agreement provides evidence that the coding framework could be applied consistently across raters, reconciliation does not eliminate the interpretive judgement involved in assigning categories to students' self-reported accounts.
	
	Third, RQ3 relies on students’ self-reported perceptions of both output correctness and their confidence in evaluating it. Moreover, our analysis characterises intervention behaviour only among interactions that students perceived as imperfect; it may therefore not capture adaptations or other forms of oversight applied to outputs that students initially rated as perfect. These measures consequently characterise perceived, rather than objectively established, output correctness and evaluation ability. We accounted for repeated observations within students using mixed-effects modelling, and the relationship between perceived minor flaws and evaluation confidence remained robust when the five-point confidence measure was modelled as ordinal. However, only five interactions were classified as \textsc{Major Hallucination Errors}. We therefore excluded these observations from inferential comparisons, and no conclusions can be drawn about evaluation confidence under severe AI failure.
	
	Finally, the study was conducted in a single first-year, design-first, low-code software engineering course at the University of Sydney. The prominence of fitness-oriented oversight and output transformation may therefore partly reflect the course's emphasis on requirements, system design, scope management, and reflective AI use. The findings should not be assumed to generalise directly to programming-intensive courses, more experienced software engineers, or professional software development settings. Nevertheless, the semester-long coursework setting provides evidence of oversight within authentic educational activities rather than isolated experimental tasks.
	
	\section{Conclusion}\label{sec:conclusion}
	As GenAI increasingly participates in the production of software engineering artefacts, students must learn not only to use these systems but also to exercise effective oversight of their outputs. This study examined how novice software engineering students enacted such oversight during authentic, semester-long coursework.
	
	Our findings show that oversight extends substantially beyond detecting AI errors. Fitness problems were identified considerably more often than correctness problems in students' reported interventions. They exercised oversight by modifying generated artefacts, refining prompts, seeking clarification, or rejecting outputs. Students' evaluation confidence was also lower when they perceived outputs as containing minor flaws, highlighting the challenge of evaluating imperfect but plausible AI-generated artefacts.
	
	Taken together, the findings show that novice oversight involves more than identifying errors in AI-generated outputs. Students also judged whether outputs were suitable for their engineering context and responded by modifying the artefact, redirecting the AI, seeking clarification, or rejecting the output. Understanding oversight in AI-assisted software engineering therefore requires attention not only to whether students recognise when an AI output is incorrect, but also to how they judge its \emph{fitness for purpose} and respond when it does not meet the needs of the engineering task.
	
	Future work should triangulate reflective journal accounts with direct interaction traces and examine whether the distinction between correctness- and fitness-oriented oversight persists across programming-intensive courses and professional software engineering contexts.
	
	\section*{Ethics} 
	This study was approved by the Human Research Ethics Committee of the University of Sydney (Approval No. HREC 2025/HE001132-substudy 022 v1a). 
	
	\section*{Data Availability} 
	The underlying student interaction records cannot be made publicly available because the conditions of the human research ethics approval restrict the dissemination of participant data to de-identified aggregate results. To support methodological transparency and verification, the complete coding framework, including code definitions and coding procedures, is provided in Section~\ref{sec:method}.
	
	\section*{Acknowledgements}
	This project was supported by an Education-Focused Academics SoTL Grant from the University of Sydney. 

	\bibliographystyle{IEEEtran}
	\bibliography{references}
	
\end{document}